\documentclass[conference, twocolumn]{IEEEtran}
\IEEEoverridecommandlockouts
\def\BibTeX{{\rm B\kern-.05em{\sc i\kern-.025em b}\kern-.08em
		T\kern-.1667em\lower.7ex\hbox{E}\kern-.125emX}}
	
\usepackage{amsmath,amssymb,amsfonts}
\usepackage{algorithm}
\usepackage{algpseudocode}
\usepackage{graphicx}
\usepackage{textcomp}
\usepackage{xcolor}

\usepackage[english]{babel}
\usepackage[utf8]{inputenc}
\usepackage[font=small]{caption}

\usepackage{placeins}
\usepackage{color}
\usepackage{mathtools}

\usepackage{parskip} 
\usepackage{xspace}
\usepackage{microtype}

\usepackage{cite} 
\usepackage{tabularx}
\usepackage{multirow}

\usepackage{gensymb} 
\usepackage{booktabs}
\usepackage{ifthen}
\usepackage{eurosym} 
\usepackage{amsbsy} 

\usepackage[mathcal]{euscript}

\usepackage{subfigure}

\usepackage{grffile}

\usepackage{icomma}
\usepackage{tabularx}
\usepackage{nicefrac} 
\usepackage{listings} 
	\lstdefinestyle{myCustomMatlabStyle}{
		tabsize=4,
		showspaces=false,
		showstringspaces=false
	}
\usepackage{transparent} 
\usepackage{import} 

\definecolor{myblue}{RGB}{62, 176, 247}
\usepackage[hidelinks]{hyperref} 

\usepackage[acronym]{glossaries} 

\makeatletter
\def\IEEElabelanchoreqn#1{\bgroup
	\def\@currentlabel{\p@equation\theequation}\relax
	\def\@currentHref{\@IEEEtheHrefequation}\label{#1}\relax
	\Hy@raisedlink{\hyper@anchorstart{\@currentHref}}\relax
	\Hy@raisedlink{\hyper@anchorend}\egroup}
\makeatother
\newcommand{\subnumberinglabel}[1]{\IEEEyesnumber
	\IEEEyessubnumber*\IEEElabelanchoreqn{#1}}
	
\newcommand{\T}{\ensuremath{^\intercal}\xspace}

\newcommand{\nk}{\ensuremath{(n|k)}\xspace}
\let\k\relax 
\newcommand{\k}{\ensuremath{(k)}\xspace}
\newcommand{\kpo}{\ensuremath{(k+1)}\xspace}

\newcommand{\Np}{\ensuremath{N_\g{p}}\xspace}
\newcommand{\Npred}{\ensuremath{N_\g{pred}}\xspace}

\newcommand{\sumnNp}{\sum_{n=0}^{\Np-1}}

\newcommand{\sumnpoNppo}{\sum_{n=1}^{\Np}}

\newcommand{\uvek}{\ensuremath{u}\xspace}

\newcommand{\uk}{\ensuremath{\uvek(k)}\xspace}

\newcommand{\ukk}[1][]{
	\ifthenelse{ \equal{#1}{} }
	{\ensuremath{\uvek(k+1)}\xspace}
	{\ensuremath{\uvek(k+{#1})}\xspace}
} 

\newcommand{\useq}{\ensuremath{\boldsymbol{\uvek}}\xspace}

\newcommand{\dvek}{\ensuremath{d}\xspace}

\newcommand{\dk}{\ensuremath{\dvek(k)}\xspace}

\newcommand{\lmon}{\ensuremath{\ell_\g{mon}}\xspace}

\newcommand{\Jmon}{\ensuremath{J_\g{mon}}\xspace}

\newcommand{\Jcomfagg}{\ensuremath{J_\g{comf,agg}}\xspace}

\newcommand{\Jserveragg}{\ensuremath{J_\g{s,agg}}\xspace}
\newcommand{\Jserverdis}{\ensuremath{J_\g{s,dis}}\xspace}

\newcommand{\Jcomfdis}{\ensuremath{J_\g{comf,dis}}\xspace}

\newcommand{\cCHP}{\ensuremath{c_\g{cur}}\xspace} 

\newcommand{\epsc}{\ensuremath{\varepsilon_\g{c}}\xspace}

\newcommand{\Cthb}{\ensuremath{C_\g{th,b}}\xspace}
\newcommand{\Cthserver}{\ensuremath{C_\g{th,s}}\xspace}
\newcommand{\Cthi}[1][]{
	\ifthenelse{ \equal{#1}{} }
	{\ensuremath{C_{\g{th},i}}\xspace}
	{\ensuremath{C_{\g{th},{#1}}}\xspace}
}        
\newcommand{\Cthj}{\ensuremath{C_{\g{th},j}}\xspace}

\newcommand{\Hab}{\ensuremath{H_\g{air,b}}\xspace}
\newcommand{\Haserver}{\ensuremath{H_\g{air,s}}\xspace}
\newcommand{\Hai}[1][]{
	\ifthenelse{ \equal{#1}{} }
	{\ensuremath{H_{\g{air},i}}\xspace}
	{\ensuremath{H_{\g{air},{#1}}}\xspace}
} 

\newcommand{\betaij}{\ensuremath{\beta_{ij}}\xspace}
\newcommand{\betaji}{\ensuremath{\beta_{ji}}\xspace}
\newcommand{\betabs}{\ensuremath{\beta_\g{bs}}\xspace}

\newcommand{\cchp}{\cCHP}

\renewcommand{\sc}{\ensuremath{s_\g{c}}\xspace}

\newcommand{\thetab}{\ensuremath{\vartheta_\g{b}}\xspace}
\newcommand{\thetabd}{\ensuremath{\dot{\vartheta}_\g{b}}\xspace}

\newcommand{\thetaserver}{\ensuremath{\vartheta_\g{s}}\xspace}
\newcommand{\thetaservert}{\ensuremath{\thetaserver(t)}\xspace}
\newcommand{\thetaserverd}{\ensuremath{\dot{\vartheta}_\g{s}}\xspace}
\newcommand{\thetaserverdt}{\ensuremath{\thetaserverd(t)}\xspace}

\newcommand{\Pgrid}{\ensuremath{P_\g{grid}}\xspace}

\newcommand{\Pchp}{\ensuremath{P_\g{chp}}\xspace}

\newcommand{\Pdem}{\ensuremath{P_\g{dem}}\xspace}

\newcommand{\Qrad}{\ensuremath{\dot{Q}_\g{rad}}\xspace}
\newcommand{\Qcool}{\ensuremath{\dot{Q}_\g{cool}}\xspace}
\newcommand{\Qcooli}[1][]{
	\ifthenelse{ \equal{#1}{} }
	{\ensuremath{\dot{Q}_{\g{cool,}i}}\xspace}
	{\ensuremath{\dot{Q}_\g{cool,{#1}}}\xspace}
}

\newcommand{\Qcoolb}{\ensuremath{\dot{Q}_{\g{cool,b}}}\xspace}
\newcommand{\Qcoolbt}{\ensuremath{\Qcoolb(t)}\xspace}
\newcommand{\Qcoolserver}{\ensuremath{\dot{Q}_{\g{cool,s}}}\xspace}
\newcommand{\Qcoolservert}{\ensuremath{\Qcoolserver(t)}\xspace}

\newcommand{\Qheat}{\ensuremath{\dot{Q}_\g{heat}}\xspace}
\newcommand{\Qheati}[1][]{
	\ifthenelse{ \equal{#1}{} }
	{\ensuremath{\dot{Q}_{\g{heat,}i}}\xspace}
	{\ensuremath{\dot{Q}_\g{heat,{#1}}}\xspace}
}

\newcommand{\Qotherb}{\ensuremath{\dot{Q}_\g{other,b}}\xspace}
\newcommand{\Qotherbt}{\ensuremath{\Qotherb(t)}\xspace}
\newcommand{\Qotherserver}{\ensuremath{\dot{Q}_\g{other,s}}\xspace}
\newcommand{\Qotherservert}{\ensuremath{\Qotherserver(t)}\xspace}
\newcommand{\Qotheri}[1][]{
	\ifthenelse{ \equal{#1}{} }
	{\ensuremath{\dot{Q}_{\g{other,}i}}\xspace}
	{\ensuremath{\dot{Q}_{\g{other,{#1}}}}\xspace}
}

\newcommand{\thetaa}{\ensuremath{\vartheta_\g{air}}\xspace}
\newcommand{\thetai}[1][]{
	\ifthenelse{ \equal{#1}{} }
	{\ensuremath{\vartheta_{i}}\xspace}
	{\ensuremath{\vartheta_{\g{#1}}}\xspace}
}

\newcommand{\thetabi}[1][]{
	\ifthenelse{ \equal{#1}{} }
	{\ensuremath{\vartheta_{\g{b},i}}\xspace}
	{\ensuremath{\vartheta_{\g{b,{#1}}}}\xspace}
}

\newcommand{\thetaid}{\ensuremath{\dot{\vartheta}_{i}}\xspace}

\newcommand{\thetaj}{\ensuremath{\vartheta_{j}}\xspace}

\newcommand{\sct}{\ensuremath{E}(t)\xspace}

\newcommand{\thetabt}{\ensuremath{\thetab(t)}\xspace}
\newcommand{\thetabdt}{\ensuremath{\thetabd(t)}\xspace}

\newcommand{\thetait}[1][]{
	\ifthenelse{ \equal{#1}{} }
	{\ensuremath{\vartheta_{i}(t)}\xspace}
	{\ensuremath{\vartheta_{#1}(t)}\xspace}
} 

\newcommand{\Pgridt}{\ensuremath{P_\g{grid}}(t)\xspace}
\newcommand{\Pchpt}{\ensuremath{P_\g{chp}}(t)\xspace}

\newcommand{\Prent}{\ensuremath{P_\g{ren}}(t)\xspace}
\newcommand{\Pdemt}{\ensuremath{P_\g{dem}}(t)\xspace}

\newcommand{\Qradt}{\ensuremath{\dot{Q}_\g{rad}}(t)\xspace}

\newcommand{\thetaat}{\ensuremath{\thetaa(t)}\xspace}
\newcommand{\Qtotit}[1][]{
	\ifthenelse{ \equal{#1}{} }
	{\ensuremath{\dot{Q}_{\g{tot,}i}}\xspace}
	{\ensuremath{\dot{Q}_{\g{tot,}{#1}}}\xspace}
} 

\newcommand{\Tsamp}{\ensuremath{T_\g{s}}\xspace}

\newcommand{\scvek}{\ensuremath{E}\xspace}
\newcommand{\sck}{\ensuremath{E_\g{}(k)}\xspace}
\newcommand{\sckpo}{\ensuremath{E_\g{}(k+1)}\xspace}
\newcommand{\scdot}{\ensuremath{\dot{E}(t)\xspace}}

\newcommand{\Pgridk}{\ensuremath{P_\g{grid}(k)}\xspace}
\newcommand{\Pchpk}{\ensuremath{P_\g{chp}(k)}\xspace}

\newcommand{\Pdemk}{\ensuremath{P_\g{dem}(k)}\xspace}

\newcommand{\Qradk}{\ensuremath{\dot{Q}_\g{rad}(k)}\xspace}

\newcommand{\thetaak}{\ensuremath{\thetaa(k)}\xspace}

\newcommand{\wmon}{\ensuremath{w_\g{mon}}\xspace}

\newcommand{\wcomf}{\ensuremath{w_\g{comf}}\xspace}

\newcommand{\wserveragg}{\ensuremath{w_\g{s,agg}}\xspace}

\newcommand{\wthi}[1][]{
	\ifthenelse{ \equal{#1}{} }
	{\ensuremath{w_{\g{th},i}}\xspace}
	{\ensuremath{w_{\g{th},{#1}}}\xspace}
}

\newcommand{\Imat}[2][]{
	\ifthenelse{ \equal{#1}{} }
	{\ensuremath{\boldsymbol{I}}\xspace}
	{\ensuremath{\boldsymbol{I}_{{#1}\times {#2}} }\xspace}
}
\newcommand{\Zeromat}[2][]{
	\ifthenelse{ \equal{#1}{} }
	{\ensuremath{\boldsymbol{0}}\xspace}
	{\ensuremath{\boldsymbol{0}_{{#1}\times {#2}} }\xspace}
}
\newcommand{\Onemat}[2][]{
	\ifthenelse{ \equal{#1}{} }
	{\ensuremath{\boldsymbol{1}}\xspace}
	{\ensuremath{\boldsymbol{1}_{{#1}\times {#2}} }\xspace}
}

\newcommand{\indHLab}{\g{agg}}
\newcommand{\xHLab}{\ensuremath{x_\indHLab}\xspace}

\newcommand{\uHLab}{\ensuremath{u_\indHLab}\xspace}
\newcommand{\dHLab}{\ensuremath{d_\indHLab}\xspace}
\newcommand{\AHLab}{\ensuremath{A_\indHLab}\xspace}
\newcommand{\BHLab}{\ensuremath{B_\indHLab}\xspace}
\newcommand{\SHLab}{\ensuremath{S_\indHLab}\xspace}

\newcommand{\indLLb}{\g{dis}}

\newcommand{\xLLb}{\ensuremath{x_\indLLb}\xspace}
\newcommand{\uLLb}{\ensuremath{u_\indLLb}\xspace}

\newcommand{\dLLb}{\ensuremath{d_\indLLb}\xspace}
\newcommand{\ALLb}{\ensuremath{A_\indLLb}\xspace}
\newcommand{\BLLb}{\ensuremath{B_\indLLb}\xspace}
\newcommand{\SLLb}{\ensuremath{S_\indLLb}\xspace}
\newcommand{\errLLb}{\ensuremath{\err_\indLLb}\xspace}

\newcommand{\Eevi}[1][]{
	\ifthenelse{ \equal{#1}{} }
	{\ensuremath{E_{\g{EV},i}}\xspace}
	{\ensuremath{E_{\g{EV,{#1}}}}\xspace}
}
\newcommand{\Eevarri}[1][]{
	\ifthenelse{ \equal{#1}{} }
	{\ensuremath{E_{\g{EV,arr},i}}\xspace}
	{\ensuremath{E_{\g{EV,arr,{#1}}}}\xspace}
}
\newcommand{\Eevdepi}[1][]{
	\ifthenelse{ \equal{#1}{} }
	{\ensuremath{E_{\g{EV,dep},i}}\xspace}
	{\ensuremath{E_{\g{EV,dep,{#1}}}}\xspace}
}
\newcommand{\Eevmini}[1][]{
	\ifthenelse{ \equal{#1}{} }
	{\ensuremath{E_{\g{EV,min},i}}\xspace}
	{\ensuremath{E_{\g{EV,min,{#1}}}}\xspace}
}

\newcommand{\Cevi}[1][]{
	\ifthenelse{ \equal{#1}{} }
	{\ensuremath{C_{\g{EV},i}}\xspace}
	{\ensuremath{C_{\g{EV,{#1}}}}\xspace}
}
\newcommand{\Cevarri}[1][]{
	\ifthenelse{ \equal{#1}{} }
	{\ensuremath{C_{\g{EV,arr},i}}\xspace}
	{\ensuremath{C_{\g{EV,arr,{#1}}}}\xspace}
}
\newcommand{\Cevdepi}[1][]{
	\ifthenelse{ \equal{#1}{} }
	{\ensuremath{C_{\g{EV,dep},i}}\xspace}
	{\ensuremath{C_{\g{EV,dep,{#1}}}}\xspace}
}

\newcommand{\Pevi}[1][]{
	\ifthenelse{ \equal{#1}{} }
	{\ensuremath{P_{\g{EV},i}}\xspace}
	{\ensuremath{P_{\g{EV,{#1}}}}\xspace}
}
\newcommand{\Pevmaxi}[1][]{
	\ifthenelse{ \equal{#1}{} }
	{\ensuremath{P_{\g{EV,max},i}}\xspace}
	{\ensuremath{P_{\g{EV,max,{#1}}}}\xspace}
}

\newcommand{\Devi}[1][]{
	\ifthenelse{ \equal{#1}{} }
	{\ensuremath{D_\g{i}}\xspace}
	{\ensuremath{D_{\g{#1}}}\xspace}
}

\newcommand{\mmpi}[1][]{
	\ifthenelse{ \equal{#1}{} }
	{\ensuremath{m_{\g{MP},i}}\xspace}
	{\ensuremath{m_{\g{MP},{#1}}}\xspace}
}

\newcommand{\Ppv}{\ensuremath{P_\g{PV}}\xspace}

\newcommand{\thetampi}[1][]{
	\ifthenelse{ \equal{#1}{} }
	{\ensuremath{\vartheta_{\g{MP},i}}\xspace}
	{\ensuremath{\vartheta_{\g{MP},{#1}}}\xspace}
}

\newcommand{\pmpi}[1][]{
	\ifthenelse{ \equal{#1}{} }
	{\ensuremath{p_{\g{MP},i}}\xspace}
	{\ensuremath{p_{\g{MP},{#1}}}\xspace}
}

\newcommand{\err}{\ensuremath{\epsilon}\xspace}
\newcommand{\erri}[1][]{
	\ifthenelse{ \equal{#1}{} }
	{\ensuremath{\epsilon_{i}}\xspace}
	{\ensuremath{\epsilon_{{#1}}}\xspace}
}	

\newcommand{\erragg}{\ensuremath{\err_\g{agg}}\xspace}
\newcommand{\errb}{\ensuremath{\epsilon_\g{b}}\xspace}

\newcommand{\errserver}{\ensuremath{\epsilon_\g{s}}\xspace}

	\newcommand{\errk}{\ensuremath{\err\k}\xspace}
	\newcommand{\errik}{\ensuremath{\erri\k}\xspace}

\newcommand{\comp}{\ensuremath{\tilde{\epsilon}}\xspace}
\newcommand{\compb}{\ensuremath{\tilde{\epsilon}_\g{b}}\xspace}
\newcommand{\comps}{\ensuremath{\tilde{\epsilon}_\g{s}}\xspace}
\newcommand{\compi}{\ensuremath{\tilde{\epsilon}_{i}}\xspace}

	\newcommand{\compk}{\ensuremath{\comp\k}\xspace}
	\newcommand{\compbk}{\ensuremath{\compb\k}\xspace}
	\newcommand{\compsk}{\ensuremath{\comps\k}\xspace}
	\newcommand{\compik}{\ensuremath{\compi\k}\xspace}

\newcommand{\compilin}{\ensuremath{\tilde{\epsilon}_{i}^{\g{lin}}}\xspace}
\newcommand{\compixgb}{\ensuremath{\tilde{\epsilon}_{i}^{\g{xgb}}}\xspace}

\newcommand{\todk}{\ensuremath{\mathrm{ToD}(k)}\xspace}
\newcommand{\doyk}{\ensuremath{\mathrm{DoY}(k)}\xspace}
\newcommand{\nhist}{\ensuremath{n_\g{hist}}\xspace}

\newcommand{\MAE}{\ensuremath{\mathrm{MAE}}\xspace}

\let\st\relax
\DeclareMathOperator*{\st}{s.t.}

\newcommand{\eg}{e.\thinspace{}g.\@\xspace}
\newcommand{\ie}{i.\thinspace{}e.\@\xspace}

\newcommand{\pagerefh}[1]{\hyperref[#1]{page~\pageref*{#1}}}
\newcommand{\secref}[1]{\hyperref[#1]{Section~\ref*{#1}}}
\newcommand{\chapref}[1]{\hyperref[#1]{Chapter~\ref*{#1}}}
\newcommand{\figref}[1]{Figure~\ref{#1}}
\newcommand{\tabref}[1]{\hyperref[#1]{Table\ \ref*{#1}}}

\newcommand{\algoref}[1]{Algorithm~\ref{#1}}
\newcommand{\g}{\mathrm}

\newcommand{\kW}{\ensuremath{\g{kW}}\xspace}
\newcommand{\kWh}{\ensuremath{\g{kWh}}\xspace}

\newcommand{\kWpK}{\ensuremath{\frac{\g{kW}}{\g{K}}}\xspace}
\newcommand{\kWpKnice}{\ensuremath{\nicefrac{\g{kW}}{\g{K}}}\xspace}
\newcommand{\kWhpK}{\ensuremath{\frac{\g{kWh}}{\g{K}}}\xspace}
\newcommand{\kWhpKnice}{\ensuremath{\nicefrac{\g{kWh}}{\g{K}}}\xspace}

\newcommand{\kWel}{\ensuremath{\g{kW}_\g{el}}\xspace}

\newcommand{\kWpeak}{\ensuremath{\g{kWp}}\xspace}

\newcommand{\degC}{\ensuremath{\degree \g{C}\xspace}}

\newcommand{\qm}{\ensuremath{\g{m}^2}\xspace}

\newcommand{\hours}  {\ensuremath{\,\g{h}}\xspace}

\newcommand\MEUR[1]{\if@EURleft\text{\euro}\,\fi#1\if@EURleft\else\,\text{\euro}\fi}

\newcommand{\Matlab}{\textsc{Matlab}\xspace}

\newcommand{\percent}{\,\%\xspace}

\newcommand{\parodis}{PARODIS\xspace}

\makeglossaries

\newacronym{bem}{BEM}{building energy management}
\newacronym{res}{RES}{renewable energy source}
\newacronym{mpc}{MPC}{Model Predictive Control}
\newacronym{moo}{MOO}{multi-objective optimization}
\newacronym{ev}{EV}{electric vehicle}
\newacronym{bps}{BPS}{building performance software}
\newacronym{ode}{ODE}{ordinary differential (or difference) equation}
\newacronym{hri}{HRI}{Honda Research Institute}
\newacronym{hvac}{HVAC}{heating, ventilation, and air conditioning}
\newacronym{chp}{CHP}{combined heat and power plant}
\newacronym{pv}{PV}{photovoltaic}
\newacronym{empc}{EMPC}{Economic Model Predictive Control}
\newacronym{der}{DER}{distributed energy resource}
\newacronym{ess}{ESS}{energy storage system}
\newacronym{tes}{TES}{thermal energy storage}
\newacronym{ocp}{OCP}{optimal control problem}
\newacronym{lp}{LP}{linear programming}
\newacronym{qp}{QP}{quadratic programming}
\newacronym{nlp}{NLP}{nonlinear programming}
\newacronym{milp}{MILP}{mixed-integer linear programming}
\newacronym{minlp}{MINLP}{mixed-integer nonlinear programming}
\newacronym{dm}{DM}{decision maker}
\newacronym{awds}{AWDS}{adaptive weight determination scheme}
\newacronym{nbi}{NBI}{normal boundary intersection}
\newacronym{chim}{CHIM}{convex hull of individual minima}
\newacronym{fpbi}{FPBI}{focus point boundary intersection}
\newacronym{cup}{CUP}{closest to Utopia point}
\newacronym{aep}{AEP}{angle to the extreme points}
\newacronym{atn}{ATN}{angle to the neighbor points}
\newacronym{ci}{CI}{carbon intensity}
\newacronym{hl}{HL}{higher level}
\newacronym{ll}{LL}{lower level}

\newacronym[\glslongpluralkey={Gaussian Processes}]{gp}{GP}{Gaussian Process}
\newacronym{bepst}{BEPST}{building energy performance simulation tool}
\newacronym{ann}{ANN}{artificial neural network}
\newacronym{rnn}{RNN}{recurrent neural network}
\newacronym{ems}{EMS}{energy management system}
\newacronym{svr}{SVR}{support vector regression}
\newacronym{dr}{DR}{demand response}
\newacronym{ga}{GA}{Genetic Algorithm}
\newacronym{rc}{RC}{Resistor-Capacitor}
\newacronym{rbf}{RBF}{radial basis function}
\newacronym{pmv}{PMV}{predicted mean vote}
\newacronym{ekf}{EKF}{extended Kalman filter}
\newacronym{ml}{ML}{machine learning}

\newacronym{sil}{SiL}{software-in-the-loop}
\newacronym{mae}{\MAE}{mean absolute error}
\newacronym{fmu}{FMU}{functional mock-up unit}
	
\glsdisablehyper

\begin{document}

\title{Regression-Based Model Error Compensation \\ for a Hierarchical MPC \\Building Energy Management System\\
}

\author{Thomas Schmitt$^{1}$, 
	Jens Engel$^{1}$, 
	Tobias Rodemann$^{1}$
	\thanks{$^{1}$Honda Research Institute Europe GmbH, 
		Offenbach, Germany. 
		E-mail: {\tt\footnotesize \{thomas.schmitt, jens.engel, tobias.rodemann\}@honda-ri.de}
	}%
}

\maketitle

\begin{abstract}
	One of the major challenges in the development of \glspl{ems} for complex buildings is accurate modeling.
	To address this, we propose an \gls{ems}, which combines a \gls{mpc} approach with data-driven model error compensation.
	The hierarchical \gls{mpc} approach consists of two layers:
	An aggregator controls the overall energy flows of the building in an aggregated perspective, while a distributor distributes heating and cooling powers to individual temperature zones.
	The controllers of both layers employ regression-based error estimation to predict and incorporate the model error.
	The proposed approach is evaluated in a \acrlong{sil} simulation using a physics-based digital twin model.
	Simulation results show the efficacy and robustness of the proposed approach.
\end{abstract}


\glsresetall
%

\section{Introduction}

The increasing penetration of \glspl{res} in the public power grid leads to a demand for intelligent \glspl{ems} for buildings. 
The most popular method for controlling \glspl{ems} is \gls{mpc}.
However, for \gls{mpc} to be effective, an appropriate model of a building's energy behavior is necessary. 

There are several approaches to building such a model, which can be categorized as white-box modeling, gray-box modeling, and black-box modeling. 
White-box models, mostly developed using \acrlongpl{bepst} such as EnergyPlus or TRNSYS, can be very accurate, but are usually too complex to be used directly in the \gls{mpc}'s \gls{ocp}. 
Grey-box models, such as state space or \gls{rc} models, are less accurate, but can be utilized well in an \gls{ocp} \cite{privara2013building}. 
Both white- and gray-box modeling of buildings is very complex and requires building-specific expert knowledge, \ie models cannot be easily transferred to other buildings \cite{privara2013building}. 
Thus, data-driven black-box modeling has experienced an increase in interest \cite{wang2019data}, \eg using \glspl{gp} or \glspl{ann}. 
While the biggest advantage is the comparatively low modeling effort, they require a large amount of data, the aqcuisition of which is again challenging \cite{maddalena2020data}. 
At the same time, including possibly known dynamics or behavior are difficult to incorporate directly and may also have to be approximated.
Therefore, a hybrid approach of these modeling paradigms is likely necessary to succeed in employing building \glspl{ems} in a larger scale in the real world. 

One option is to replace (a part of) the building's model by a \textit{data-driven surrogate model}. 
In \cite{stoffel2022combining,tang2022data}, a machine learning model is trained with simulation data from a physics-based model. 
Then, the machine learning model is included in the \gls{mpc}'s \gls{ocp}.  
Data-driven surrogate models are also frequently used for real-world buildings. 
In \cite{jain2020neuropt}, \glspl{ann} are trained with historical data from a test building located at the University of L'Aquila, Italy to predict both energy consumption and temperature development. 
The \glspl{ann} are then utilized as the sole model in the \gls{mpc}. 
In \cite{huang2015hybrid}, \acrlongpl{rnn} are used to approximate a nonlinear thermal model of an airport check-in hall. 
The check-in hall's temperature is then controlled using \gls{mpc} to both follow a reference trajectory and not violate comfort boundaries by solving a linear \gls{ocp}. 
However, \glspl{ann} can also be used as part of the objective function, instead of replacing model equations in the constraints. 
In \cite{ferreira2012neural}, \gls{rbf}-based \glspl{ann} are used to approximate both the thermal dynamics and the occupant comfort
for 4 university office rooms.
For more examples of data-driven control approaches,  
the reader is referred to the review \cite{kathirgamanathan2021data}. 
Notably, only very few studies consider multi-zone buildings.

A second option for a hybrid model approach is a \textit{data-driven error estimator} (or \textit{residual estimator}). 
Here, the goal is not to replace a part of the gray-box model, but to reduce the model error by augmenting it with a residual value, estimated by a data-driven regression model. 
However, applications in the building sector are sparse. 
In \cite{massagray2018hybrid}, a physics-based model of a single-office building in Stuttgart, developed in TRNSYS and \Matlab, is first simplified to a \gls{rc} gray-box model. 
Then, a \gls{gp} model is trained to predict the error of the \gls{rc} model, using simulation data from the physics-based model as ground truth.  
However, it was not applied to any control purposes. 
Applications of error estimators in combination with \gls{mpc} can be found in different areas. 
In \cite{jain2020bayesrace}, \glspl{gp} are used to learn the model error for an autonomous racing car. 
Training data is received from simulation without \gls{mpc}. 
The \glspl{gp} are explicitly used in the \gls{ocp} as part of the model dynamics. 
In \cite{yoo2017event}, a \gls{rbf}-based disturbance estimator for a nonholonomic robot is used for event-triggered \gls{mpc}. 
The disturbance is assumed to be dependent on the system state and control input only, and could thus be interpreted as a model error. 
In \cite{carron2019data}, a \gls{gp} based error estimation is combined with an \acrlong{ekf} to achieve offset-free tracking of a 6 degrees of freedom robotic arm. 

In this work, we use a hierarchical setup for the \gls{mpc} of the energy system of a medium-sized office building in Offenbach, Germany.
An aggregator is used to control the total energy flows, which are then allocated to the 9 individual temperature zones by a distributor. 
Gray-box state space models are used on both levels. 
A physics-based digital twin serves as a surrogate model of the actual building. 
To compensate the model errors of both the aggregator and the distributor, we train two regression-based error estimators. 
As features, only signals which are easily obtainable both online and offline are used. 
Training data is derived from a \gls{sil} simulation of the digital twin with real-world measurement data. 
The main contributions are 
the development of the data-driven estimators for a multi-zone building using a digital twin and real-world measurement data, 
and their application for error compensation in a hierarchical \gls{mpc} approach. 

The rest of the paper is structured as follows.
The building itself as well as its digital twin and the simplified gray-box models are described in \secref{sec:modeling}. 
The hierarchical \gls{mpc} setup is explained in \secref{sec:control_approach}. 
The data-driven error estimators and their training process is discussed in \secref{sec:error_comp_methodology}. 
The successful error compensation by combining the error estimators with the hierarchical \gls{mpc} approach is shown by long-term simulation results in \secref{sec:simulation_results}.
Finally, we conclude with a discussion on the impacts and necessary further steps in \secref{sec:conclusion}.

\section{Building Models} 
\label{sec:modeling}

In this section, we will first give a brief description of the actual building. 
Then, we will explain the different models used in this study, \ie 
1) the digital twin, 
2) a state-space model with only a single temperature zone used by the aggregator. 
and 3) a state-space model of the 9 temperature zones used by the distributor.   

\subsection{Building Description}

The building used in this study is a medium-sized company building located in Offenbach, Germany. 
It has a footprint of approx. $13,\,000\,\qm$ and can be separated into 9 different temperature zones, which include offices, halls, some workshops and, as a peculiarity, an emissions lab. 
Besides the connection to the public power grid, the main energy sources are 
a gas-fired \gls{chp} for co-production of electricity and heat with 199\,\kWel, 
a fairly large \gls{pv} plant with 750\kWpeak, which serve an average load demand of approx. $250\,\kW$.
It further has gas-fired heating boilers and an electric \gls{hvac} system. 
A stationary second-life battery with a capacity of $98\,\kWh$ can be used as electric storage. 

\subsection{Digital Twin}
\label{subsec:modeling_digitaltwin}
A Modelica-based simulation model implemented in SimulationX is used as a digital twin \cite{unger2012green}. 
It covers 
the 9 different temperature zones, 
their couplings, 
heat losses to both the ambient air and the ground, 
internal heat gains from electrical consumption and occupants, 
and the above mentioned energy producers and consumers, including various constraints on the power production. 
The \gls{chp} has a minimal power output of 50\percent, below which it cannot be modulated.
Furthermore, its power-up and and power-down times, as well as nonlinear efficiencies are considered.
The SimulationX model uses historic measurement data for 
the ambient air temperature, 
the electric power demand (per zone), 
solar irradiation, 
and \gls{pv} power production. 

\subsection{Aggregator Model}
As discussed in the introduction, the physics-based digital twin is not suited to be used in an \gls{ocp}. 
Thus, we use a simplified state space model representing the most important entities. 
Note that the hierarchization, \ie the use of an aggregator and a distributor, is done to ensure the scalability of the control approach. 
This also allows the integration of additional components, \eg charging stations for electric vehicles \cite{engel2022hierarchical}. 
Of the total 9 temperature zones, 7 are aggregated as a single 'building zone' with an average temperature \thetab (in \degC). 
The remaining 2 zones refer to server rooms and are aggregated with an average 'server zone' temperature \thetaserver (in \degC). 
The stationary battery's stored energy \scvek (in \kWh) completes the state vector \xHLab. 
The inputs \uHLab to the system consist of 
the grid power \Pgrid, 
the (electrical) \gls{chp} power \Pchp, 
the gas heating power \Qrad, 
and the \gls{hvac} cooling power \Qcool.

As disturbances \dHLab,  
\gls{pv} power \Ppv, 
the building's electrical power demand \Pdem, 
the ambient air temperature \thetaa (in \degC), 
(constant) losses to the ground \Qotherb,
and (constant) internal heatings \Qotherserver
are considered.
All powers are given in \kW. 
The time-continuous state space model is then given by 
\begin{IEEEeqnarray}{rcl}\label{eq:MGmodel_time}
	\begin{bmatrix}
		\scdot \\ \thetabdt \\ \thetaserverdt
	\end{bmatrix} & = &
	\begin{bmatrix}
		0 & 0 & 0 \\ 
		0 & -\frac{\Hab - \betabs}{\Cthb} & \frac{\betabs}{\Cthb} \\
		0 & \frac{\betabs}{\Cthserver} & -\frac{\Haserver - \betabs}{\Cthserver} 
	\end{bmatrix} \cdot
	\begin{bmatrix}
		\sct \\ \thetabt \\ \thetaservert
	\end{bmatrix} \ldots \nonumber \\ 
	\IEEEeqnarraymulticol{3}{l}{
		+\> \!\!   
		\begin{bmatrix}
			1 & 1	 					 & 0 				  & \frac{1}{\epsc}  & \frac{1}{\epsc}  \\
			0 & \frac{1}{\cchp\cdot\Cthb} & \frac{1}{\Cthb}	  & \frac{1}{\Cthb} 	 & 0 \\
			0 & 0 & 0 & 0 & \frac{1}{\Cthserver}
		\end{bmatrix} \cdot
		\begin{bmatrix}
			\Pgridt \\ \Pchpt \\ \Qradt \\ \Qcoolbt \\ \Qcoolservert
		\end{bmatrix} \ldots
	}\nonumber \\
	\IEEEeqnarraymulticol{3}{l}{
		+\> \!\!
		\begin{bmatrix}
			1 & 1 & 0 & 0 & 0 \\
			0 & 0 & \frac{\Hab}{\Cthb} & \frac{1}{\Cthb} & 0 \\
			0 & 0 & \frac{\Haserver}{\Cthserver} & 0 & \frac{1}{\Cthserver}
		\end{bmatrix} \cdot
		\begin{bmatrix}
			\Prent \\ \Pdemt \\ \thetaat \\ \Qotherbt \\ \Qotherservert
		\end{bmatrix},
	} \IEEEeqnarraynumspace
\end{IEEEeqnarray}
where 
\Cthb and \Cthserver are the thermal capacities of the building and server zone, respectively in \kWhpK, 
\Hab and \Haserver are the heat transfer coefficients to the ambient air of the building and the server zone in \kWpK, respectively, 
\betabs is the heat transfer coefficient between the two zones in \kWpK and 
\cchp is the ratio of the \gls{chp}'s electrical to thermal power. 
Numerical values are given in \tabref{tab:num_values}. 

In the following, we only use its discretized state space form.  
Furthermore, we respect the model errors 
for the building and the server zone temperatures, \ie 
\begin{IEEEeqnarray}{rCl}\label{eq:HLab_dis_ss}
	\xHLab(k+1) &=& \AHLab(\Tsamp)\xHLab(k) + \BHLab(\Tsamp)\uHLab(k) \ldots \IEEEnonumber \\ 
	&& +\> \SHLab(\Tsamp)\dHLab(k) + \erragg\k
	\IEEEeqnarraynumspace 
	\label{eq:HLabc_agg_ss}
\end{IEEEeqnarray}
with $\erragg\k = \begin{bmatrix} 0 & \errb\k & \errserver\k \end{bmatrix}\T$ and \Tsamp being the sampling rate in $\g{h}$. 
Note that we can respect the model error $\erragg\k$ only in the discretized form since it has to be estimated from discretely sampled data points. 
For more details on the modeling itself, the reader is referred to \cite{schmitt2022multi}.

\subsection{Distributor Model}
The distributor models the 9 temperature zones individually, while neglecting the electrical part of the aggregator model. 
The temperature $\thetai$ of a single zone $i$ can be described by
\begin{IEEEeqnarray}{rCl}
	\thetaid(t) &=& 
	\frac{1}{\Cthi} \left( \Qheati(t) + \Qcooli(t) + \Qotheri(t) \right)\ldots \IEEEnonumber  
	\IEEEeqnarraynumspace  
	\\
	&& 
	-\> \sum_{j \neq i} \frac{\betaij}{\Cthi} \left( \thetai(t) - \thetaj(t) \right) \ldots \IEEEnonumber 
	\\
	&& 
	- \frac{\Hai}{\Cthi} \left( \thetai(t) - \thetaat \right) 
\end{IEEEeqnarray} 
with
$\Cthi$ being the thermal capacity of zone $i$ in $\kWhpK$, 
\betaij the heat transfer coefficient between zones $i$ and $j$ in $\kWpK$, 
$\Hai$ the heat transfer coefficient between zone $i$ and the outside air in $\kWpK$, 
and $\Qheati$ and $\Qcooli$ in \kW the heating and cooling powers allocated to zone $i$. 
$\Qotheri$ is an uncontrollable disturbance, which is assumed constant and either represents heat losses to the ground (for the building zones 1-7) or internal heat gains (for the server zones 8 and 9). 

Using the 9 $\thetai$ as states \xLLb, 
\Qheati and \Qcooli as inputs \uLLb,  
\Qotheri as disturbances \dLLb, and
again discrete model errors 
$\errLLb\k = 
\begin{bmatrix}
	\erri[1]\k & \ldots & \erri[9]\k
\end{bmatrix}\T$,  
they are expressed as the discrete state space model 
\begin{IEEEeqnarray}{rCl}\label{eq:LLb_dis_ss}
	\xLLb(k+1) &=& \ALLb(\Tsamp)\xLLb\k + \BLLb(\Tsamp)\uLLb\k \ldots \IEEEnonumber \\
				&& +\> \SLLb(\Tsamp)\dLLb\k + \errLLb\k, \IEEEeqnarraynumspace
\end{IEEEeqnarray}	
where \ALLb is the system matrix, \BLLb the input matrix and \SLLb the disturbance matrix. 
Again, \Tsamp denotes the sampling time and the numerical values are given in \tabref{tab:num_values}.
For brevity, the reader is referred to \cite{schmitt2022multi} for more details on the state space model. 

\begin{table}[htb]
	\begin{center}
		\caption{
			Numerical values of the building parameters.  
			Note that 
			$\Cthb = \sum_{i=1}^{7}\Cthi$, 
			$\Cthserver = \Cthi[8] + \Cthi[9]$, 
			$\Hab = \sum_{i=1}^{7}\Hai$,
			$\Haserver = \Hai[8] + \Hai[9]$,
			$\betabs = \beta_{29} + \beta_{58} + \beta_{68}$, 
			$\betaij = \betaji$, 
			and all other \betaij not listed below are zero, \eg $\beta_{12} = 0$. 
		}
		\label{tab:num_values}
		\begin{tabular}{lrclrclr}
			\toprule
			&   in $\kWhpKnice$ & &  &   in $\kWpKnice$  &  & &   in $\kWpKnice$ \\ \midrule[0.2pt] 
			$\Cthi[1]$ &	230.88 &  &$\Hai[1]$  & 3.69	& &	$\beta_{29}$	& 48.40 		\\
			$\Cthi[2]$ &	476.29 & & $\Hai[2]$  & 9.82	& &	$\beta_{34}$	& 345.60		\\
			$\Cthi[3]$ &	214.27 & & $\Hai[3]$  & 3.65	& &	$\beta_{56}$	& 1100.48	\\
			$\Cthi[4]$ &	103.68 & & $\Hai[4]$  & 2.79	& &	$\beta_{58}$	& 23.40		\\
			$\Cthi[5]$ &	330.14 & & $\Hai[5]$  & 4.79	& &	$\beta_{68}$	& 8.00			\\
			$\Cthi[6]$ &	330.14 & & $\Hai[6]$  & 6.19	& &		& 					\\
			$\Cthi[7]$ &	99.456 & & $\Hai[7]$  & 3.19	& &		& 					\\
			$\Cthi[8]$ &	2.40   & & $\Hai[8]$  & 0.03	& &		& 					\\
			$\Cthi[9]$ &	4.80   & & $\Hai[9]$  & 0.04	& &		& 					\\
			\bottomrule
		\end{tabular}
	\end{center}
\end{table}


\section{Control Approach}
\label{sec:control_approach}
In this section, we describe the \glspl{ocp} solved by the \gls{mpc} on both the aggregator and distributor level.

\subsection{Aggregator Control}
\label{subsec:control_approach_aggregator}

The aggregator's goal is to regulate the building temperatures while minimizing the monetary costs. 
This is expressed as a weighted sum of multiple cost functions. 
First, for the building zone temperature, the so-called comfort costs 
\begin{IEEEeqnarray}{rCl}
	\Jcomfagg(k) & = & \sumnpoNppo  \left( \thetab\nk - 22\degC \right)^2, \label{eq:lcomf} 
\end{IEEEeqnarray}
apply. 
The notation $\thetab\nk$ refers to the value for $\thetab(k+n)$ predicted at time step $k$. 
\Npred is the number of steps in the prediction horizon. 
Second, the monetary costs are expressed as 
\begin{IEEEeqnarray}{rcl}
	\Jmon(k) & = &\!\sumnNp\!\lmon\!\left(\!\Pgrid\nk,\Pchp\nk,\Qheat\nk,\Tsamp\!\right) \IEEEeqnarraynumspace \label{eq:Jmon} 
\end{IEEEeqnarray}
where \lmon describes the costs arising from gas usage and buying (selling) electrical energy from (to) the public grid. 
Note that we consider German industry pricing, in which different prices for buying and selling as well as high peak costs apply. 
Details on both numerical values and how \Jmon can be reformulated using an epigraph formulation, which results in a linear programming problem, can be found in \cite[pp.\,24]{schmitt2022multi}. 
Third, the server zone is only kept within an acceptable temperature range by 
\begin{IEEEeqnarray}{rCCl}
	\Jserveragg\k & = & \sumnpoNppo & 
	  \max\left( 15\degC - \thetaserver\nk,~0 \right) \ldots \IEEEnonumber \\
	  &&& +\> \max\left( \thetaserver\nk - 21\degC,~0  \right).  
	\IEEEeqnarraynumspace \label{eq:Jserveragg} 
\end{IEEEeqnarray}
The input constraints are given by 
\begin{IEEEeqnarray}{rCCCl}
	\subnumberinglabel{eq:constraints_HL_inputs_all}  
	-1000\,\kW  & \leq & 			\Pgridk	 & \leq & 1000\,\kW  \label{eq:constraints_HL_input_Pgrid}\\
	0  			& \leq & 			\Pchpk	 & \leq & 199\,\kW,  \label{eq:constraints_HL_input_Pchp}\\
	0  			& \leq & 			\Qradk   & \leq & 1500\,\kW, \label{eq:constraints_HL_input_Qrad}\\ 
	-1353\,\kW 	& \leq & 			\Qcoolb\k  & \leq & 0, 		 \label{eq:constraints_HL_input_Qcoolb}\\
	-197\,\kW 	& \leq & 			\Qcoolserver\k  & \leq & 0.  \label{eq:constraints_HL_input_Qcoolserver}
\end{IEEEeqnarray}
The state constraints are given by
\begin{IEEEeqnarray}{rCCCl}
	\subnumberinglabel{eq:constraints_HL_states_all}  	
	0.15\cdot 98\,\kWh  & \leq & \sck 	& \leq & 0.85\cdot 98\,\kWh, \IEEEeqnarraynumspace  \label{eq:constraints_HL_battery_Cbat}\\
	-32.9\,\kW		   & \leq & \frac{\sckpo-\sck}{\Tsamp} & \leq & 32.9\,\kW. \label{eq:constraints_HL_battery_Pbat}
\end{IEEEeqnarray}
Note that \thetab and \thetaserver are unconstrained to avoid infeasibilities in the later co-simulation without error compensation. 
Both are only regulated due to the respective cost functions. 

Together, the aggregator's \gls{ocp} is described by 
\begin{IEEEeqnarray}{cCl}
	\IEEEyesnumber* \IEEEyessubnumber*
	\min_{\useq_\g{\indHLab}} & ~ & \wcomf \cdot \Jcomfagg\k + \wmon \cdot \Jmon\k \ldots \IEEEnonumber \\
					   && +\> \wserveragg \cdot \Jserveragg\k, 
	\IEEEeqnarraynumspace \label{eq:opt_prob_HLab_mon_ind_comf} 
	\\
	\st & & \eqref{eq:HLabc_agg_ss},~\eqref{eq:constraints_HL_inputs_all} ~\forall\, n = 0 \ldots \Npred -1,   \\
	&& \eqref{eq:constraints_HL_states_all} ~\forall\, n = 1 \ldots \Npred,   
\end{IEEEeqnarray} 
with 
$\useq_\g{\indHLab} = \left( \uHLab(0|k),\, \ldots\,,\,  \uHLab(\Npred-1|k)\right)$ 
being the sequence of control inputs, and a prediction horizon of $\Npred = 48$ steps of $\Tsamp=0.5\hours$ each, \ie 1 day in total. 
The time step notation $(k)$ and $(k+1)$ in \eqref{eq:HLabc_agg_ss}, \eqref{eq:constraints_HL_inputs_all} and \eqref{eq:constraints_HL_states_all} are to be read as $(n|k)$ and $(n+1|k)$, respectively.

Usually, the weights are chosen such that a reasonable compromise is determined \cite{schmitt2020application}. 
Alternatively, multi-objective optimization can be used \cite{schmitt2020multi,schmitt2022incorporating}, since the aggregator's \gls{ocp} is always solvable quickly enough due to the hierarchization. 
However, here we choose $\wcomf = 0.99, \wmon = 0.01, \wserveragg = 0.99$ to ensure that the controller tries to achieve $\thetab = 22\degC$ and $15\degC \leq \thetaserver \leq 21\degC$ at all times. 
This simplifies the evaluation of the error compensation later on.
Note that in the actual implementation, additional slack variables are used due to the reformulation of \Jmon and of the $\max$-terms of \Jserveragg. 

Since we want to assess the compensation of the model error, we simulate with no prediction error. 
Namely, we assume perfect predictions for the PV power, the building's load and the ambient air temperature. 
For an assessment of the influence of real predictions for the facility under study, the reader is referred to \cite{wang2019automated} and \cite{schmitt2021cost}. 

\subsection{Distributor Control}
\label{subsec:control_approach_distributor}

In the distributor, the total heating and cooling powers determined by the aggregator are split (distributed) between the individual zones. 
To this end, we use individual weights $\sum_{i=1}^{9}\wthi = 1$ proportional to the thermal capacities, \ie 
\begin{IEEEeqnarray}{rCl}
	\wthi & = & \frac{ \Cthi}{\sum_{j=1}^{9} \Cthj} ~\forall\,i=1, \ldots, 9. \label{eq:wthi}
\end{IEEEeqnarray}
The temperature goals are the same as in the aggregator, \ie we punish temperature deviations from $22\degC$ in the 7 building zones by 
\begin{IEEEeqnarray}{rCl}
	\Jcomfdis(k) & = & \sumnpoNppo 
						\sum_{i=1}^{7} \wthi \cdot \left( \thetai\nk - 22\degC \right)^2.   
						\IEEEeqnarraynumspace \label{eq:Jcomfdis}
\end{IEEEeqnarray}
For the 2 server zones, the same temperature range applies as in the aggregator. 
Outside of these, we punish temperature deviations by 
\begin{IEEEeqnarray}{rCCl}
	\Jserverdis(k) & = & \sumnpoNppo  \sum_{i=8}^{9} 
								& 
								\Big( 
								\max\left( 15\degC - \thetai\nk,~0 \right) 
								\ldots \IEEEnonumber \IEEEeqnarraynumspace 
								\\
					&&& +\> \max\left( \thetai\nk - 21\degC,~0  \right)  
								\Big).  
					\IEEEeqnarraynumspace \label{eq:Jserverdis}
\end{IEEEeqnarray}
The inputs are subject to box constraints which stem from the building's internal infrastructure, 
\begin{IEEEeqnarray}{rCCCl}
	\subnumberinglabel{eq:constraints_LL_inputs_box}  
	0 	& \leq & \Qheati\k ~\forall\,i=1, \ldots, 7  & \leq & 893.95\,\kW,  		 \IEEEeqnarraynumspace \label{eq:constraints_HL_input_box_Qheat1to7}\\
	0 	& \leq & \Qheati\k ~\forall\,i=8, \ldots, 9  & \leq & 0,  		 \IEEEeqnarraynumspace \label{eq:constraints_HL_input_box_Qheat8to9}\\
	-800\,\kW 	& \leq & \sum_{i \in \{1, 2, 3, 4, 7\}}	\Qcooli\k & \leq & 0, 		 \label{eq:constraints_HL_input_box_Qcool12347}\\
	-330\,\kW 	& \leq & 			\Qcooli[5]\k + \Qcooli[6]\k  & \leq & 0, 		 \label{eq:constraints_HL_input_box_Qcool56}\\
	-53\,\kW 	& \leq & 			\Qcooli[8]\k  & \leq & 0, 		 \label{eq:constraints_HL_input_box_Qcool8}\\
	-144\,\kW 	& \leq & 			\Qcooli[9]\k  & \leq & 0.  		 \label{eq:constraints_HL_input_box_Qcool9}
\end{IEEEeqnarray}
Note that the server zones 8 and 9 have no heating systems, since they have to be cooled all the time. 
Furthermore, the total powers are constrained by the powers allocated by the aggregator, 
\begin{IEEEeqnarray}{rCl}
	\subnumberinglabel{eq:constraints_LL_inputs_equality}  
	\sum_{1}^{7}\Qheati\k 	& = & 			\Qheat\k + \frac{\Pchp\k}{\cchp}, 		 \label{eq:constraints_HL_input_eq_Qcoolb}\\
	\sum_{1}^{7}\Qcooli\k 	& = & 			\Qcoolb\k, 		 \label{eq:constraints_HL_input_eq_Qcoolb}\\
	\sum_{8}^{9}\Qcooli\k 	& = & 			\Qcoolserver\k.  		 \label{eq:constraints_HL_input_eq_Qcoolserver}
\end{IEEEeqnarray}
As in the aggregator, the zone temperatures have no hard constraints to avoid infeasibilities in the co-simulation with no error compensation. 

Together, the distributor's \gls{ocp} is described by 
\begin{IEEEeqnarray}{cCl} 
	\IEEEyesnumber* \IEEEyessubnumber*
	\min_{\useq_\g{\indLLb}} & ~ & \Jcomfdis\k + \Jserverdis\k 
	\IEEEeqnarraynumspace 
	\\
	\st & & \eqref{eq:LLb_dis_ss},~ \eqref{eq:constraints_LL_inputs_box},~\eqref{eq:constraints_LL_inputs_equality} ~\forall\, n = 0 \ldots \Npred-1
	\label{eq:opt_prob_Distributor} 
\end{IEEEeqnarray} 
with 
$\useq_\g{\indLLb} = \left( \uLLb(0|k),\, \ldots\,,\,  \uLLb(\Npred-1|k)\right)$ 
being the sequence of control inputs, 
and the same prediction horizon as in the aggregator. 
Again, the time step notation $(k)$ and $(k+1)$ in \eqref{eq:LLb_dis_ss}, \eqref{eq:constraints_LL_inputs_box} and \eqref{eq:constraints_LL_inputs_equality} are to be read as $(n|k)$ and $(n+1|k)$, respectively. 

\section{Error Compensation Methodology}
\label{sec:error_comp_methodology}
As previously described, the control approach is aware of a model error $\errk$ in both the aggregator and distributor.
We aim to perform error compensation, \ie we want to find an estimator $\compk$ that can approximate this error, such that ${\compk \approx \errk}$.
Incorporating the estimator to approximate the model error should improve control performance.
We use machine learning regression models to build these estimators.
We train 9 estimators $\compik, i = 1, \dots, 9$, \ie one for each temperature zone in the distributor.
The estimators in the aggregator are the weighted sum of the individual zone estimators, \ie 
\begin{IEEEeqnarray}{rCl}\IEEEyesnumber \IEEEyessubnumber*
	\compbk & = & \frac{\sum_{i=1}^{7} \wthi \cdot \compik}{\sum_{i=1}^{7} \wthi},  \\
	\compsk & = & \frac{\sum_{i=8}^{9} \wthi \cdot \compik}{\sum_{i=8}^{9} \wthi}.
\end{IEEEeqnarray}
This is analogous to $\thetab$ and $\thetaserver$ being the weighted averages of the individual zone temperatures $\thetai$. 
The estimators will predict the model error for only one time step at a time, \ie $\Np$ separate predictions will be made to calculate $\compi\nk$ over the horizon $n = 0, \ldots, \Np-1$ at each time step $k$.

\subsection{Feature selection}
\label{subsec:feature_selection}
The first step to training a regression model is feature selection.
The target variable (\ie labels) of the regression model are the measured model errors.
In principle, these can be calculated as the difference between observed state and predicted state, \ie $\errk = x\kpo - x(1|k)$.
Generally, the resulting difference may also include prediction errors of the disturbances.
This can be circumvented by recalculating $x(1|k)$ using the state space model and measurements of disturbances $\dk$ and inputs $\uk$.
For selecting the features, we want to consider that the resulting estimators should be easily (re)trainable and employable.
This means that we should only use features that are readily available and both measurable and predictable.
Therefore, all disturbances of the MPC controller are good candidate features, as they are both measurable and predictable in the case of the proposed EMS.
From a brief correlation analysis between measured errors $\errk$ and measured disturbances (omitted for brevity), we deducted that a set of 5 features should provide good a basis for training, \ie
\begin{enumerate}
	\item the current ambient temperature $\thetaak$, for unaccounted heat flows to/from the environment (\eg inaccurate heat transfer coefficients; warm/cold air from ventilation)
	\item the past values of the ambient temperature $\thetaa(k-1), \dots, \thetaa(k-\nhist)$, for heat diffusion from other zones
	\item the total building load $\Pdemk$, as electrical consumption is transformed into heat and is correlated to  occupant behavior
	\item the time of day in $\hours$, \ie $\todk: \mathbb{N} \rightarrow [0, 24)$, for regular occupant behavior and ventilation schedule
	\item the day of the year $\doyk: \mathbb{N} \rightarrow \{1, \dots, 365\}$, for mapping seasonal effects. 
\end{enumerate}

\subsection{Models}
Based on these features, we propose two candidate regression models as estimators: 1) A linear regression model and 2) an XGBoost regression model.
XGBoost (eXtreme Gradient Boosting) is an open-source software library that provides an efficient and effective implementation of the gradient boosting framework for machine learning \cite{chen2016xgboost}. 
It uses gradient boosting \cite{friedman2001greedy} to improve the performance of decision trees, which can be used for both regression and classification problems.

For the first estimator, we propose the linear model
\begin{IEEEeqnarray}{rCl}
	\compilin\k &=& 
	 +\> \gamma_{i,1} \sin\Big(\frac{2\pi}{24}\todk\Big) + \gamma_{i,2} \cos\Big(\frac{2\pi}{24}\todk\Big) \nonumber \\
	&& +\> \delta_{i,1} \sin\Big(\frac{2\pi}{365}\doyk\Big) + \delta_{i,2} \cos\Big(\frac{2\pi}{365}\doyk\Big) \nonumber \\
	&& +\> \alpha_i \Pdemk + \sum_{j=0}^{\nhist} \beta_{i,j} \thetaa(k-j) + \kappa_i \IEEEeqnarraynumspace
\end{IEEEeqnarray}
for each zone $i$, where $\nhist = 2$. The parameters $\alpha_i, \beta_{i,j}, \gamma_{i,l}, \delta_{i,l}, \kappa_i$ are fitted through least-squares regression. 
The features $\todk$ and $\doyk$ are transformed using a cyclical transformation to normalize them uniquely to values between -1 and 1, preserving the cyclical nature of day time and seasons.

For the second estimator, $\compixgb\k$, we train an XGBoost regressor for each zone $i$.
For this estimator, we use all aforementioned features and $\nhist = 2$.
Contrary to the linear model, we do not apply a cyclical transformation to the time features, as this is not needed with tree-based regression models and can actually be detrimental.

\subsection{Training and evaluation}
\label{subsec:training_and_evaluation}
To generate the training data for training the estimators, we use the digital twin model described in \secref{subsec:modeling_digitaltwin}.
We simulate a full calendar year using the digital twin in a \gls{sil} setup together with the described control approach and without compensation, \ie $\compk = 0$.
In this setup, the digital twin running in SimulationX is connected through an FMU (functional mock-up unit) to a Python bridge, linking it to the MPC controller implemented in \Matlab using the \parodis framework \cite{schmitt2021parodis}.
At each discrete time step $k$, the controller receives the updated system states from the digital twin model, determines the control input $\uk$ and applies it to the digital twin.
For the simulation in the digital twin and the predictions for the \gls{mpc} controller, we use measurement data for the weather and electrical demands collected for the year 2021 at the Honda R\&D facility in Offenbach, Germany. 
During the simulation, we collect both the states predicted by the \gls{mpc} as well as the realized (\ie measured) states.
From these we calculate the model error $\errk$ for training.
In a real world setting, one would use a baseline controller to run in parallel to a controller with an estimator pre-trained in a digital twin setting, to be able to calculate raw model errors to retrain estimators on new data.

One of the main error sources between digital twin models of buildings and reality, next to occupant behavior, are the estimated heat capacities of the building zones \cite{tian2018review}.
Therefore, we benchmark the robustness of our proposed error compensation against this error, by creating additional simulation scenarios, where we change the heat capacities in the model of the MPC, while keeping them the same in the digital twin model.
Overall, we examine four scenarios, namely
\begin{enumerate}
	\item heat capacities in \gls{mpc} model are exact, 
	\item heat capacities in \gls{mpc} are 50\percent of digital twin, 
	\item heat capacities in \gls{mpc} are 150\percent of digital twin, 
	\item total heat capacity of the building is exact, individual capacities are shifted randomly according to \algoref{algo:capacity_shift}.
\end{enumerate}
To train the estimators, we use a $70/30$ train-test split on the collected data, and use scikit-learn \cite{pedregosa2011sklearn} to fit the linear regressor as well as the scikit-learn interface of the XGBoost Python library for training the XGBoost estimator, respectively.

\begin{algorithm}[htbp]
	\caption{Shifting of capacities between adjacent zones}\label{algo:capacity_shift}
	\begin{algorithmic}
		\ForAll{zone coupling pairs $(i, j)$}
		\State Draw capacity shift fraction $p_\g{shift} \gets \mathcal{U}[0, 0.5]$
		\State Randomly decide shift direction $d_\g{shift} \gets \{-1, 1\}$
		\State Calculate $C_\g{shift} = \min (C_i, C_j) \cdot p_\g{shift} \cdot d_\g{shift}$
		\State Update $C_i = C_i + C_\g{shift}$
		\State Update $C_j = C_j - C_\g{shift}$
		\EndFor
	\end{algorithmic}
\end{algorithm}

\tabref{tab:compensator_performance} shows the performance of the trained estimators on the training and test data sets in terms of \gls{mae} of the residual model error $\compik - \errik$.
We calculate the overall \gls{mae} as the weighted sum of the \gls{mae} of each temperature zone over the data set, \ie
\begin{IEEEeqnarray}{rCl}
	\MAE & = & \sum_{i=1}^9 w_i \Bigg( \frac{1}{N} \sum_{k=0}^{N-1} |\compik - \errik| \Bigg).
\end{IEEEeqnarray}
The linear estimator shows fair performance on both data sets.
The XGBoost estimator shows very good performance on the training set and similar performance on the test set.
This suggests that the estimator is not overfitting.

\begin{table}[htbp]
	\caption{Performance of trained estimator models under different scenarios in terms of \acrfull{mae}. The baseline performance column refers to the measured model error in the baseline simulations without error compensation.}
	\centering
	\def\arraystretch{1.2}%
	\begin{tabular}{llcccc}
		\hline
		\multicolumn{1}{l}{\multirow{2}{*}{Estimator}} & \multicolumn{1}{l}{\multirow{2}{*}{Scenario}} & \multicolumn{ 4}{c}{\gls{mae} in $10^{-3}\,\g{K}$} \\ \cline{ 3- 6}
		& & Baseline & Train & Test & SiL \\ \midrule
		Linear  &          2021 &                38.766 &            15.853 & 15.836    &       16.904 \\
		&          2022 &                42.137 &               --- &              --- &           20.461 \\ \midrule
		XGBoost &          2021 &                38.766 &             4.846 &            5.379 &            7.440 \\
		&          2022 &                42.137 &               --- &              --- &           15.571 \\
		&      2021 50\% &                76.596 &             7.893 &            8.718 &           12.820 \\
		&     2021 150\% &                27.544 &             5.616 &            6.291 &            9.532 \\
		&  2021 shifted &                39.473 &             6.073 &            6.653 &            8.521 \\
		\bottomrule
	\end{tabular}
	\label{tab:compensator_performance}
\end{table}

\section{Simulation Results}
\label{sec:simulation_results}
\begin{figure}[tbp]
	\centering
	\includegraphics[width=\columnwidth]{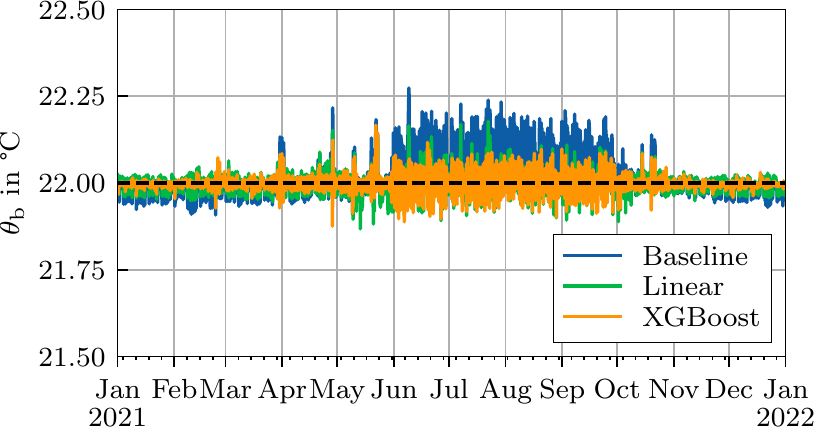}
	\caption{Comparison of overall building temperature $\thetab$ for the year 2021 between no error compensation (Baseline), the linear estimator model (Linear) and the XGBoost estimator model (XGBoost).}
	\label{fig:overall_building_temp_comparison_2021}
\end{figure}

To evaluate the performance of the proposed error compensation approach, we applied the trained estimators in the previously described \gls{sil} simulation.
First, we simulated the baseline 2021 simulation with active compensation.
\figref{fig:overall_building_temp_comparison_2021} shows the resulting average temperature of the building zone for the baseline case compared to linear compensation and XGBoost compensation.
This shows that the control performance regarding the comfort costs in the aggregator (\ie deviation from the setpoint of $22\degC$) is significantly improved with active error compensation.

\begin{figure}[tbp]
	\centering
	\includegraphics[width=\columnwidth]{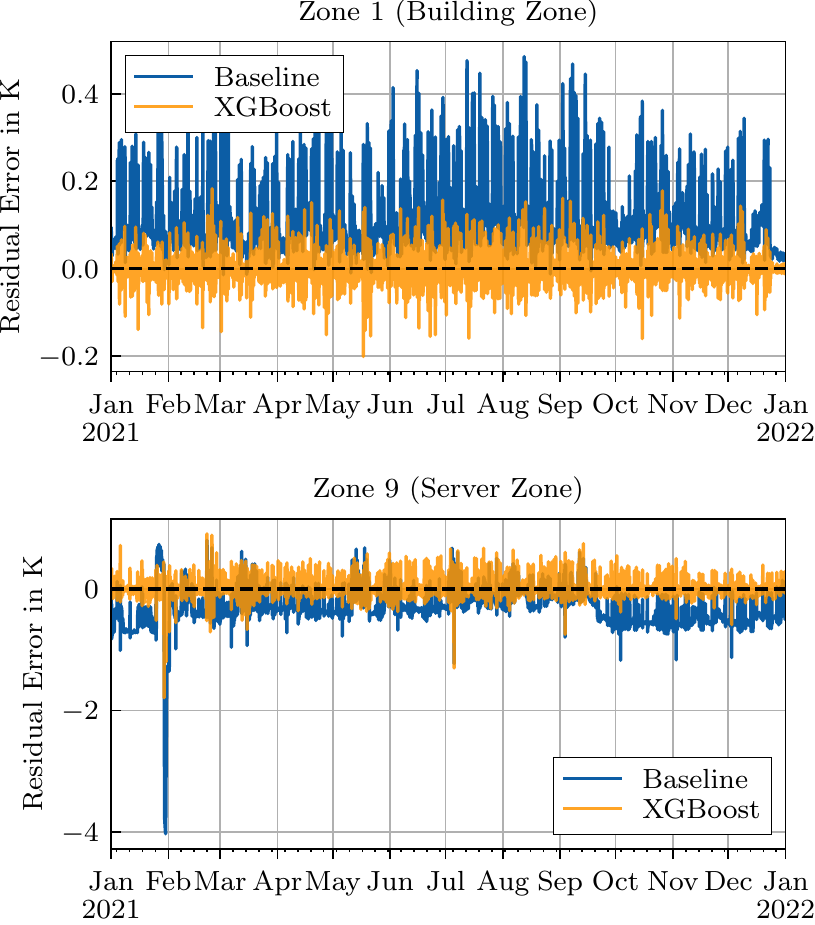}
	\caption{Comparison of the residual model error in zones 1 and 9 without compensation (Baseline) and with the XGBoost estimator model for the year 2021.}
	\label{fig:comparison_zone_1_9_residuals}
\end{figure}

\figref{fig:comparison_zone_1_9_residuals} shows the residual errors, \ie left over model error, for zones 1 and 9 in the baseline case compared to the active compensation using the XGBoost estimator.
This suggests that the estimator manages to approximate the actual model error also in the SiL simulation.
This is confirmed by looking at the overall \gls{mae} of the residual errors, as shown in \tabref{tab:compensator_performance}.
The \gls{mae} of both the linear estimator and the XGBoost estimator are significantly lower than in the baseline case, with XGBoost clearly outperforming the linear model.
The linear estimator reduces the \gls{mae} by 56\percent, the XGBoost estimator by 80\percent.
Both the linear and XGBoost estimator have a slightly decreased performance in the \gls{sil} setting.

\begin{figure}[tbp]
	\centering
	\includegraphics[width=\columnwidth]{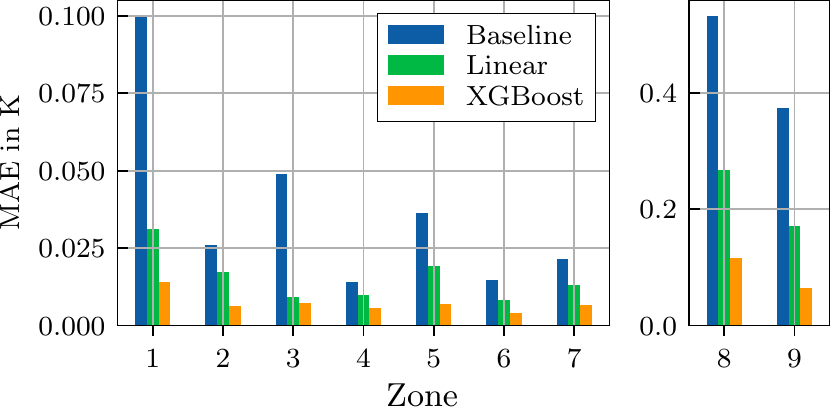}
	\caption{\Acrfull{mae} of the residual model error in all temperature zones with no error compensation (Baseline), with the linear estimator model (Linear), and with the XGBoost estimator model (XGBoost).}
	\label{fig:comparison_metric_zones}
\end{figure}

The performance of the estimators is further illustrated in \figref{fig:comparison_metric_zones}, where the \gls{mae} of each of the 9 temperature zones is shown and compared between the three cases.
This again shows that both estimators manage to approximate the model error well in all zones with XGBoost exhibiting best performance.

To evaluate the generalizability of the proposed approach, we tested the performance of the estimators trained on 2021 data in a SiL simulation with unseen data of the year 2022.
This way, we can implicitly test robustness against changing occupant behavior and weather.
We simulated the year 2022 analogously to the 2021 base scenario, \ie with exact capacities in the controller, both without compensation and with each of the estimators.
The results are again shown in \tabref{tab:compensator_performance}.
Both estimators exhibit decreased performance on 2022 data.
The linear estimator reduces the \gls{mae} by 51\percent, the XGBoost estimator by 63\percent.
The results suggest that periodical retraining of these data-driven estimators may be necessary.
For example, the facility under study experienced significant change in occupant behavior between 2021 and 2022 due to COVID-19 policies.

As motivated in \secref{subsec:training_and_evaluation}, we want to test the robustness of the approach against errors in estimated heat capacities in the controller.
We therefore simulated the three described scenarios, \ie 1) heat capacities at 50\percent, 2) heat capacities at 150\percent, and 3) randomly shifted heat capacities in the MPC, with error compensation.
We only simulated with the better performing XGBoost estimator.
The results are again shown in \tabref{tab:compensator_performance}.
In all cases, the estimators manage to significantly reduce the residual model error, while yielding only slightly worse performance than with exact heat capacities.
Overall, the results suggest that the approach is reasonably robust against this type of error.

\newpage
\section{Conclusion \& Outlook}
\label{sec:conclusion}
We have shown that our proposed data-driven error compensation approach can significantly reduce the residual model error between the proposed hierarchical \gls{mpc} controller and a digital twin building model in a \gls{sil} simulation.
We have proposed two simple regression-based error estimator models, which achieve an error reduction of up to 56\percent (linear model) and 80\percent (XGBoost) in a baseline full calendar year simulation.
We have shown that the proposed approach is robust against model errors of heat capacities in the controller.
Furthermore, we have shown that the regression-based estimators generalize reasonably well by applying the estimators trained on 2021 measurement data to a simulation based on 2022 measurement data.
Despite significant change in occupant behavior between 2021 and 2022, both estimators exhibit good performance.

\begin{figure}[htbp]
	\centering
	\includegraphics[width=\columnwidth]{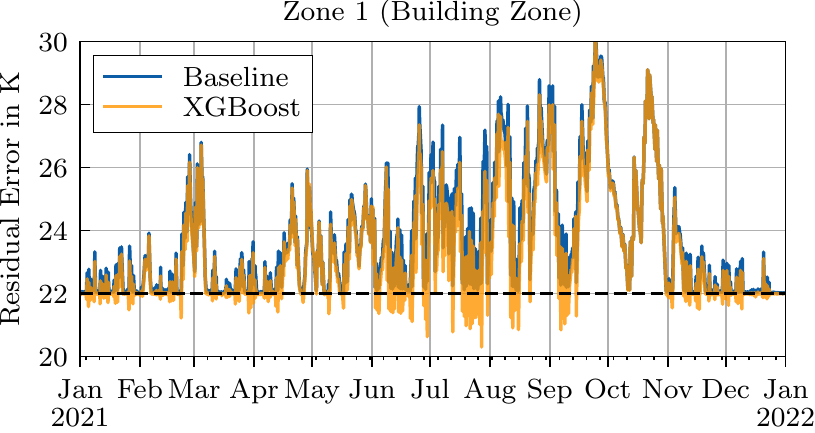}
	\caption{Temperature in zone 1 in the year 2021 without error compensation (Baseline) and with compensation using the XGBoost estimator (XGBoost).}
	\label{fig:outlook_zone_1_temperature}
\end{figure}

While the proposed error compensation approach achieves significant model error reduction in all zones and improved control performance of the overall building temperature, as shown in \figref{fig:overall_building_temp_comparison_2021}, the control performance in individual zones is still lacking.
This is illustrated in \figref{fig:outlook_zone_1_temperature}, where the temperature of zone 1 over the course of the year 2021 is shown, with and without error compensation.
The control performance is only marginally better in the case with active error compensation.
This is due to the structure of the hierarchical control approach: The aggregator derives a heating and cooling budget by considering the weighted sum of the individual zone errors of the distributor.
Thereby, positive and negative components cancel out.
In turn, not enough heating and cooling budget is allocated for compensation in the distributor.
This problem could be resolved in future work by extending the control scheme by introducing additional communication between the two layers.

For this paper, we have used data from a full calendar year for training.
However, further investigation could be conducted to understand how the amount of training data relates to the performance of the error compensation, to determine how much data is needed to (re)train compensators in a real life setting.

\FloatBarrier
\newcommand{\BIBdecl}{\setlength{\itemsep}{-6pt}}
\bibliographystyle{IEEEtran}
\bibliography{common/lib_error_comp}
	
\end{document}